\documentstyle[preprint,aps,epsf]{revtex}
\def \x {{\bf x}}
\def \y {{\bf y}}
\def \A {{\bf A}}
\def \B {{\bf B}}
\def \D {{\bf D}}
\def \E {{\bf E}}
\def \r {{\bf r}}
\def \e {{\rm e}}

\begin{document}

\title{
%\hfill\parbox[t]{2in}{\rm\small JLAB-TH-96-xx}
%\vskip 1.3cm
On the Dirac Structure of Confinement}

\author{Adam P. Szczepaniak$^1$ and Eric S. Swanson$^{1,2}$}

\address{$^1$\ 
   Department of Physics,
   North Carolina State University,
   Raleigh, North Carolina 27695-8202\\
	 $^2$\ 
   Thomas Jefferson National Laboratory, 
   12000 Jefferson Avenue, 
   Newport News, VA 23606} 

\maketitle

\begin{abstract}
The Dirac structure of confinement is shown to be of timelike-vector nature 
in the heavy quark limit of QCD. This stands in contradiction 
with the phenomenological success of the Dirac scalar confining potential. 
A resolution is achieved through the demonstration that an effective scalar
interaction is dynamically generated by nonperturbative mixing between ordinary
and hybrid $Q \bar Q$ states. The resolution depends crucially on the collective
nature of the gluonic degrees of freedom.  This implies that dynamical 
gluonic effects are vital when attempting to incorporate fine structure in
models of the $Q \bar Q$ interaction.

\end{abstract}
\date{October, 1996}
\narrowtext

\section{Introduction} 

Although it has been postulated for more than 30 years, 
the phenomenon of quark confinement remains an enigmatic feature of QCD.
Quenched lattice gauge theory and heavy quark phenomenology indicate that the
static ($m_q >> \Lambda_{QCD}$) long range potential should be linear with a 
slope 
of $b \approx 0.18\, {\rm GeV}^2$. The order $1/m_q^2$ quark-antiquark 
long range  spin-dependent (SD) structure 
has also been studied. Comparison with spin splittings in the $J/\psi$ and
$\Upsilon$ spectra\cite{schnitzer} indicate that the spin-dependence 
 can only arise
from the nonrelativistic reduction of a scalar current quark-antiquark 
interaction. This picture is also supported by calculations of the long range
spin-dependent effective potentials on the lattice\cite{michael}.

Unfortunately, little analytical progress has been made on this problem.
The framework for most investigations on this subject was provided by 
the Wilson loop approach 
of Eichten and Feinberg\cite{EF} who extended the analysis of the 
spin-independent potentials by Brown and Weisberger~\cite{BW}. 
The standard parameterization for the long range SD quark-antiquark 
interaction introduced in Ref.~\cite{EF} is given by

\begin{eqnarray}
V_{SD}(r) &=& \left( {\bbox{\sigma}_q \cdot {\bf L}_q \over 4 m_q^2} -
{\bbox{\sigma}_{\bar q} \cdot {\bf L}_{\bar q} \over 4 m_{\bar q}^2} \right) \left( {1\over r}
{d \epsilon \over d r} + {2 \over r} {d V_1 \over d r} \right) + 
\left( {\bbox{\sigma}_{\bar q} \cdot {\bf L}_q \over 2 m_q m_{\bar q}} - 
        {\bbox{\sigma}_q \cdot {\bf L}_{\bar q} \over 2 m_q m_{\bar q}} \right) 
	\left( {1 \over r} {d V_2 \over d r} \right) \nonumber \\
&& 
+ {1 \over 12 m_q m_{\bar q}}\Big( 3 \bbox{\sigma}_q \cdot \hat {\bf r} \, 
 \bbox{\sigma}_{\bar q}\cdot \hat {\bf r} -   \bbox{\sigma}_q\cdot  
 \bbox{\sigma}_{\bar q} \Big) V_3(r) 
+ {1 \over 12 m_q m_{\bar q}} \bbox{\sigma}_q \cdot \bbox{\sigma}_{\bar q} 
V_4(r). \label{VSD}
\end{eqnarray}
Here $\epsilon=\epsilon(r)$ is the static potential,  
$r=|{\bf r}|= |{\bf r}_q - {\bf r}_{\bar q}|$ is the ${\bar Q Q}$ separation 
 and the $V_i=V_i(r)$ are determined by  
 electric and magnetic 
field insertions on quark lines in the Wilson loop expectation value 
(explicit expressions are given in Section II.C).
As shown by Gromes \cite{G} covariance under Lorentz transformation 
leads to a  constraint
exists between the SD potentials, 

\begin{equation}
\epsilon(r) = V_2(r) - V_1(r).\label{grom}
\end{equation}
Other more or less fundamental 
relations were also derived~\cite{EF,G1,B2}.  

In a model approach, ${Q\bar Q}$ interactions are typically derived from 
a nonrelativistic reduction of a relativistic current-current interaction. 
As far as long range potentials are concerned only time-like vector or 
scalar currents are relevant~\cite{Grev}.
Performing a nonrelativistic reduction of a vector--vector interaction yields
$ V_1=0$, $V_2 = \epsilon$, $V_3 = \epsilon'/r - \epsilon''$, and $V_4 = 
2 \nabla^2\epsilon$. Alternatively, the reduction of a scalar 
interaction yields

\begin{equation}
V_1 = -\epsilon,\;\;  V_2=V_3=V_4 = 0.\label{scal}
\end{equation}
It is the alternation in sign of the combination $V_1 + V_2$ between 
vector and scalar currents
which, through the analysis of the heavy quarkonia spectrum, enabled 
Schnitzer to identify the scalar interaction as the likely
structure for confinement~\cite{Sch}. 

Calculations\cite{mbp,nora} of the SD potentials based on sophisticated 
models of the Wilson loop typically 
yield results which are in agreement with Eq.~(\ref{scal}). In particular the 
 ``minimal area law" model for the Wilson loop, a simple extension of the 
strong coupling limit of lattice QCD,    
 leads to a picture which is very close to the classical flux tube model
of Buchm{\"u}ller\cite{B}. He noted that if one assumes that the
chromoelectric field is confined within a tube between the $\bar Q Q$ pair then 
the magnetic field generated by the flux tube vanishes in the individual 
quark rest frames 
and the only contribution to the fine structure comes from the 
kinematical Thomas
precession. This results in a SD structure which is identical 
to that of Eq.~(\ref{scal}). 

A consistent
picture of a QCD-generated effective scalar confinement interaction appears
to be emerging. It is therefore disconcerting that attempts to build  
Hamiltonian-based models of QCD\cite{ham} seem to require vector confinement. 
For example, we have found \cite{ssjc,ss} that it is impossible to construct
a stable BCS-like vacuum of QCD when scalar confinement is assumed. This is
problematical if one wishes to dynamically generate  constituent quark masses.
Furthermore, attempts at modeling chiral pions will be hindered by the
explicit lack of chiral symmetry in a scalar interaction.
These observations
appear to stand in contradiction to the well established 
scalar confinement hypothesis.

In the following we resolve this issue by first performing a Foldy-Wouthuysen 
reduction of the full Coulomb gauge Hamiltonian of QCD. This immediately
establishes that the Dirac structure of confinement for heavy quarks is of 
a timelike-vector
nature.  It is commonly stated in the literature that vector confinement is
ruled out by its spin-dependent structure. We wish to stress that one must be
careful in this judgement. In particular, the spin-dependent interactions 
which are generated by QCD are more complicated than those given by the
simple nonrelativistic reduction of an effective long range interaction. 
Indeed, we shall demonstrate that 
the scalar character of the spin splittings 
in heavy quarkonia is dynamically generated through effective 
interactions which crucially depend on the collective nature of the gluonic
degrees of freedom.

\section{Heavy Quark Expansion of $H_{QCD}$}

Our starting point is the Coulomb gauge QCD Hamiltonian\cite{TDL}

\begin{equation}
H_{QCD} = \int d\x \psi^\dagger(\x)\left[-i\bbox{\alpha}\cdot \nabla + 
\beta m\right]\psi(\x) + H_{YM} + V_C 
- g \int d\x \psi^\dagger(\x) \bbox{\alpha}\cdot {\bf A}(\x) \psi(\x) 
\label{QCD}
\end{equation}

\noindent
where

\begin{equation}
V_C = {1\over 2} g^2\int d\x d{\bf y} \, {\cal J}^{-1} \rho^a(\x) V^{ab}({\bf x}, 
{\bf y}; A) {\cal J} \rho^b({\bf y}),\label{vc}
\end{equation}

\begin{equation}
V_{ab}(\x,\y;A) = \langle \x,a |(\nabla\cdot {\cal D})^{-1} 
(-\nabla^2) (\nabla\cdot {\cal D})^{-1} | \y,b \rangle, \label{vab}
\end{equation}

\noindent
and

\begin{equation}
H_{YM} =  {1\over 2}\int d\x \left[ {\cal J}^{-1}\bbox{\Pi}(\x) {\cal J} 
\bbox{\Pi}(\x) + {\bf B}^2(\x) \right].
\end{equation}

\noindent
The degrees of freedom are the transverse gluon field $\A=\A^a{\rm T}^a$,
its conjugate momentum  $\bbox{\Pi}=\bbox{\Pi}^a {\rm T}^a$, and the quark field in the 
Coulomb gauge. The Faddeev-Popov determinant is written as
${\cal J} = \mbox{Det}[\nabla\cdot {\cal D}]$, with ${\cal D}^{ab} = \nabla\delta^{ab} 
- g f^{abc}\A^c$ being the  covariant derivative in the adjoint representation,
and the magnetic field is given by 
${\bf B}^a = \nabla\times \A^a + g f^{abc}\A^b \times \A^c$. The static 
interaction $V_C$ is the nonabelian analog of the Coulomb potential. It 
involves the full QCD 
color charge density which has both quark and gluon components,
\begin{equation}
\rho^a(\x) = \psi^\dagger(\x) {\rm T}^a \psi(\x) + f^{abc} \A^b(\x) \cdot
\bbox{\Pi}^c(\x).
\label{rho}
\end{equation} 

The most salient feature of the Coulomb gauge Hamiltonian is that all of
the degrees of freedom are physical. This makes it especially useful for
identifying the physical mechanisms which drive the spin splittings in
heavy quarkonia.

\subsection{The Foldy-Wouthuysen Transformation}

We proceed by performing a Foldy-Wouthuysen transformation on the QCD
Hamiltonian. This is done in complete analogy to the quantum mechanical
case where an operator is constructed which removes the interactions
between upper and lower components of the quark wave function order by order
in the inverse quark mass\cite{F}, 
except that the unitary transformation is now constructed in
Fock space.
 The resulting Hamiltonian  is given by 
 
\begin{eqnarray}
H_{QCD} \to H_{FW} &=& \int d {\bf x} \left( m_q h^\dagger({\bf x}) h({\bf x}) - 
m_{\bar q} \chi^\dagger({\bf x}) \chi({\bf x})\right) + H_{YM} + V_C + H_1 + H_2 + 
\ldots,  \\
H_1 &=& {1 \over 2m_q} \int d{\bf x} h^\dagger({\bf x}) \left(\D^2 - g \bbox{\sigma} \cdot 
{\bf B} \right) h({\bf x}) - ( h \rightarrow \chi; m_q \rightarrow m_{\bar q}), \\
H_2 &=& {1 \over 8 m_q^2} \int d{\bf x} h^\dagger({\bf x}) 
 g \bbox{\sigma}\cdot [{\bf E}, \times {\bf D}] 
h({\bf x}) + ( h \rightarrow \chi; m_q \rightarrow m_{\bar q}). 
\end{eqnarray}

\noindent
In this expression  $h = (1 + \beta) \psi/2$ and $\chi = (1 - \beta) \psi/2$ 
denote the upper and lower components of the quark wave function and correspond
to the annihilation and 
creation operators of the heavy quark and antiquark respectively. 
The ellipsis denotes terms which are either of $O(1/m^3)$ or are 
spin-independent at order 
$1/m^2$. Finally  $\D = i \nabla + g \A$ is the covariant derivative in the 
fundamental representation. The electric field 
contains both transverse and longitudinal components, ${\bf E}^a = - \bbox{\Pi}^a + \E^a_\parallel$, where

\begin{equation}
\E^a_\parallel =  -\nabla A^a_0 - g \nabla \nabla^{-2} f^{abc} \A^b\cdot
\nabla A_0^c
\label{e16}
\end{equation}
and

\begin{equation}
A^a_0(\x) = g \int d\y V^{ab}(\x,\y;A)\rho^b(\y).  
\end{equation}

\subsection{The Static Potential}

To leading order in the quark mass the Hamiltonian describes two static, 
noninteracting quarks. 
At ${\cal O}(m^0)$ the Hamiltonian reduces to $H_0 = H_{YM} + V_C$.
The eigenstates of $H_0$ may be labeled by
 the quark and antiquark coordinates and by an index which classifies the adiabatic state
of the gluonic degrees of freedom, $n_r$,
\begin{equation}
H_0 |n_r; \r_q \r_{\bar q} \rangle = \epsilon_n(r) |n_r; \r_q 
\r_{\bar q}\rangle.
\end{equation}
Note that we have made explicit the
dependence of the gluonic degrees of freedom on the position of the quarks, $r$.

The corresponding eigenenergies, $\epsilon_n(r)$  may be identified with the 
Wilson loop potentials calculated on 
the lattice. Thus for example, $\epsilon_0(r)$ is the Coulomb plus linear 
potential seen long ago \cite{latt}. Static hybrid states are collectively denoted
$\vert n_r;\r_q \r_{\bar q}\rangle$ with  $n_r  \ne 0$. In  recent studies \cite{pm} 
 the  lowest lying 
adiabatic hybrid potential, $\epsilon_1(r)$ has been evaluated. 

While both $H_{YM}$ and $V_C$ may contribute to the linearly rising potential energy seen 
on the lattice, it is clear that the quarks may
only interact with the flux tube via the nonabelian Coulomb interaction. Therefore the Dirac 
structure of confinement corresponds to $\gamma_0\otimes\gamma_0$  from the product of color 
 charge densities (see Eqs.~(\ref{vc}) and ~(\ref{rho})). 
As stressed in the
Introduction, this appears to be at odds with 20 years of quark model 
phenomenology. Since the phenomenology is based on spin splittings, it 
will be instructive to examine the $1/m^2$ perturbative 
corrections to the static potential.

\subsection{Spin-dependent Potentials}

The spin-dependent first order correction to the static potential is given by

\begin{equation}
\delta\epsilon_n^{(1)}(r) = {g \over 8 m_q^2} \langle n_r; \r_q \r_{\bar q}\vert
 \int d\x h^\dagger(\x) \bbox{\sigma}\cdot [{\bf E}, \times {\bf D}] h(\x) 
\vert n_r;\r_q \r_{\bar q}\rangle + ( h \rightarrow \chi; 
\, m_q \rightarrow m_{\bar q}).\label{e18}
\end{equation}

\noindent
The order $1/m$ term is not considered because $\D^2$ is not spin-dependent and
the matrix element of $\bbox{\sigma}\cdot\B$ vanishes. Eq.~(\ref{e18}) may be
simplified considerably as follows. Since we are interested in spin-dependent
terms only, the covariant derivative may be replaced with the ordinary
derivative,  $i\nabla$, and the
electric field may be replaced by  $-\nabla A^0$. Contracting the fermion field
operators and using Eq.~(\ref{e16}) yields 

\begin{equation}
\delta\epsilon_n^{(1)}(r) \sim {-i g^2 \over 8 m_q^2} \epsilon^{ijk} \sigma_q^i 
\langle n_r \vert \nabla_{r_q}^j({\rm T}^a V^{ab}(\r_q,\r_{\bar q};A){\rm T}^b) 
\vert n_{r} \rangle \nabla_{r_q}^k + (h \rightarrow \chi; m_q \rightarrow 
m_{\bar q}; \sigma_q \rightarrow -\sigma_{\bar q}),
\end{equation}
where the matrix elements are over
gluonic degrees of freedom only. 
The approximation sign is meant to serve as a reminder that the 
equality holds for spin dependent terms only.
The above equation may be simplified further by using the following relation

\begin{equation}
\langle n_r \vert (\nabla_{r_q}^j g^2 {\rm T}^a 
V^{ab}(\r_q,\r_{\bar q};A) {\rm T}^b)
\vert n_r\rangle = - \nabla_{r_q}^j \epsilon_n(r).
\end{equation}
The physical content of this relationship is simply the statement that $H_{YM}$
does not explicitly depend on the quark positions. The minus sign is due to 
contracting the antiquark operators. 
Quark line contraction also yields tadpole
terms which vanish in the color singlet background and self energy 
diagrams which are subsumed into the 
leading order, ${\cal O}(m)$, Hamiltonian. 
Making the appropriate substitution 
yields the standard classical plus Thomas precession spin orbit interaction

\begin{equation}
\delta \epsilon_n^{(1)} = 
\left( {\bbox{\sigma}_q \cdot {\bf L}_q \over 4 m_q^2} -
{\bbox{\sigma}_{\bar q} \cdot {\bf L}_{\bar q} \over 4 m_{\bar q}^2} \right)
 {1\over r}
{d \epsilon_n \over d r}.
\end{equation}

\noindent
Thus the first term in Eq.~(\ref{VSD}), generalized to any adiabatic potential and therefore true for both 
ordinary and hybrid $\bar Q Q$ states, 
is reproduced. 
At second order in perturbation theory the SD corrections are given by

\begin{equation}
\delta\epsilon_n^{(2)}(r) = \sum_{m\ne n} {\left| \langle n_r;\r_q \r_{\bar q}| H_1
 |m_r;\r_q \r_{\bar q}\rangle \right|^2 \over \epsilon_n(r) - \epsilon_m(r)}. \label{sec}
\end{equation}

\noindent
In this expression there are two terms which correspond to the application of
the magnetic field operator twice on a single quark or antiquark line. These 
matrix 
elements are not spin dependent since the product of the Pauli matrices 
collapses to unity plus a single Pauli matrix. Alternatively, the case where 
the magnetic fields
act on different quark lines is nontrivial and yields

\begin{eqnarray}
&& \delta\epsilon^{(2)}_n\vert_{BB} =  {g^2 \over 4 m_q m_{\bar q}} \sigma_q^i 
\sigma_{\bar q}^j \times \nonumber \\
&& \Bigg( \sum_{m\ne n} {\langle n_r; \r_q \r_{\bar q}\vert \int d\x 
h^\dagger(\x) B^i(\x) 
h(\x) \vert m_r; \r_q \r_{\bar q}\rangle \, \langle m_r; \r_q \r_{\bar q}\vert 
\int d\y \chi^\dagger(\y) B^j(\y)
\chi(\y) \vert n_r; \r_q \r_{\bar q}\rangle \over \epsilon_n(r) - \epsilon_m(r)} 
+ \nonumber \\ 
&& \qquad \qquad (h \leftrightarrow\chi) \Bigg). \label{ee}
\end{eqnarray}

\noindent
If we define ($g^2$ times) the term in brackets as  
$ \left[ (\hat r^i \hat r^j - {1\over 3}\delta^{ij}) V_3  + {1\over 3} 
\delta^{ij} V_4 \right]$ then the third and fourth terms in the expression of 
the spin-dependent potential in Eq.~(\ref{VSD}) follow. If time-dependent 
fields are considered Eq.~(\ref{ee}) may be rewritten as follows

\begin{equation}
\left[ (\hat r^i \hat r^j - {1\over 3}\delta^{ij}) V_3  + {1\over 3} 
\delta^{ij} V_4 \right] = \lim_{T \to \infty}  g^2  
\int_{-T/2}^{T/2} dt dt' \left({1\over T}\right) \langle n_r; \r_q \r_{\bar q} \vert
B^i({\bf r}_q,t) B^j({\bf r}_{\bar q}, t') \vert n_r; \r_q \r_{\bar q} \rangle.
\end{equation}

\noindent
Setting $n_r = 0$ in this expression yields a result which agrees with that of
Eichten and Feinberg\footnote{Their expressions are in terms of Wilson loops and
therefore project onto the ground state as $T$ goes to infinity}.

There are four terms in Eq.~(\ref{sec}) which contribute to $V_1$. 
These involve the
application of the magnetic field and the covariant derivative on the same quark
line, 

\begin{eqnarray}
&&{-g \over 4 m_q^2} \Bigg( \sum_{m\ne n} {\langle n_r; \r_q \r_{\bar q}\vert \int d\x 
h^\dagger(\x) 
\D^2 h(\x) \vert m_r; \r_q \r_{\bar q}\rangle \, \langle m_r; \r_q \r_{\bar q}\vert 
\int d\y h^\dagger(\y) 
{\bbox \sigma}\cdot \B(\y) h(\y)_ \vert n_r; \r_q \r_{\bar q}\rangle \over 
\epsilon_n(r) - \epsilon_m(r)} + \nonumber \\
&& H.c. + (h \rightarrow \chi; m_q \rightarrow m_{\bar q}) \Bigg).
\label{xxx}
\end{eqnarray}

\noindent
Using the following relationship (which holds for the spin dependent terms only) 

\begin{equation}
{d \over dt} h^\dagger(\x) \D^2 h(\x) \sim -2 i g h^\dagger(\x) {\bf E}(\x) 
\cdot \nabla h(\x), \label{rel}
\end{equation}

\noindent
allows one to replace the first factor in Eq.~(\ref{xxx}) with one involving the
electric field.  
Performing the time integral and contracting the fermion field operators 
results in 

\begin{equation}
\delta\epsilon_n^{(2)}\vert_{V_1} = {g^2 \over 2 m_q^2} \sigma_q^j \left(
\sum_{m\ne n} {\langle n_r\vert E^i(\r_q) \vert m_r\rangle \, 
\langle m_r\vert B^j(\r_q)
\vert n_r \rangle \over (\epsilon_n(r) - \epsilon_m(r))^2}  + 
H.c. \right) \nabla_{r_q}^i
+ (h \rightarrow \chi; m_q \rightarrow m_{\bar q}; \sigma_q \rightarrow 
-\sigma_{\bar q}).
\end{equation}

\noindent
If $g^2$ times the term in brackets is defined as 

\begin{equation}
-i{d V_1^{(n)}(r)\over dr} { r^k \over r} \epsilon^{ijk}
\end{equation}

\noindent
then this yields a spin orbit interaction in agreement with the second term in 
Eq.~(\ref{VSD}).
Alternatively, if the time integral is retained after substituting 
Eq.~(\ref{rel}) in Eq.~(22) and the sum over 
energy denominators is also represented by another time integral, 
then the expression 
for $V_1$ can be cast into the following form 
\begin{equation}
\hat {\bf r} {d V_1 \over d r } = \lim_{T \to \infty} { i g^2 \over 2} 
\int_{-T/2}^{T/2} dt dt' \left({t' - t \over T}\right) \langle n_r; \r_q 
\r_{\bar q}\vert {\bf B}({\bf r}_q,t) \times {\bf E}({\bf r}_q, t') 
\vert n_r; \r_q \r_{\bar q}\rangle \label{v1ef}
\end{equation}

\noindent
which agrees with the result of Eichten and Feinberg~\cite{EF}.

The remaining contributions to the second order energy correction correspond
to an interaction between the quark and antiquark with both electric and magnetic
operators at the vertices. Similar manipulations as for $V_1$
yield the portion of $V_{SD}$ proportional to $V_2$ once 
the following definition is made:

\begin{equation}
-i{dV^{(n)}_2(r) \over dr} {r^k \over r} \epsilon^{ijk} = g^2\left( \sum_m 
{\langle n_r\vert B^i(\r_q) \vert m_r\rangle \, 
\langle m_r\vert E^j(\r_{\bar q})
\vert n_r\rangle \over (\epsilon_n(r) - \epsilon_m(r))^2} + H.c. 
\right),
\end{equation}

\noindent which is equivalent to 

\begin{equation}
\hat {\bf r} {d V_2 \over d r } = \lim_{T \to \infty} {i g^2 \over 2} 
\int_{-T/2}^{T/2} dt dt' \left({t' - t \over T}\right) \langle n_r; \r_q \r_{\bar q}
\vert {\bf B}({\bf r}_{\bar q},t) \times {\bf E}({\bf r}_q, t') \vert
n_r; \r_q \r_{\bar q} \rangle. \label{v2ef}
\end{equation}

The Faddeev-Popov determinants have not been carried through the calculations
shown above. They may easily be restored without changing any of the
results.
The time integral representations of the potentials shall prove convenient for 
subsequent calculations; however they are rather opaque.
Alternatively, the application of the Foldy-Wouthuysen transformation to the 
Coulomb gauge QCD 
Hamiltonian shows that these may be simply interpreted as nonperturbative mixing 
with hybrid states. 
This makes it clear that it is possible for
nonperturbative dynamical physics to generate an effective spin-dependent
interaction which mimics a scalar interaction. 
Actually demonstrating this requires that the matrix elements be evaluated. In 
the next section we
propose to do this with the Flux Tube Model of Isgur and Paton.

\section{Model Evaluation of the Spin-dependent Potentials}

Before proceeding to a model evaluation of the matrix elements we note that 
it is possible to make some general statements on their expected
properties. The results of lattice gauge theory make it clear that a 
flux tube-like configuration of glue exists between static quarks. If one
thinks of this as a localized object with an infinite number of degrees of
freedom, then it is apparent that $V_2$ must evaluate to zero. This is because
the electric field operator creates a  local excitation in
the flux tube at position $\r_q$ (cf., Eq.~(\ref{v2ef})). This must then be de-excited at 
$\r_{\bar q}$ 
by the magnetic field operator. 
However, the two operators become decorrelated because infinitely many 
degrees of freedom intervene. 
Similarly, the long
range portions of  $V_3$ and
$V_4$ both vanish. Thus, by Gromes' relation (Eq.~(\ref{grom})),  the only nonzero long
range interaction must be given by $V_1 = - \epsilon$. This is precisely 
the situation required for ``scalar" confinement. It is therefore entirely plausible
that an effective scalar confinement is generated by nonperturbative mixing 
with hybrids. Furthermore,  the structure of the spin-dependent terms depends 
crucially on the nature of the ground state gluonic degrees of 
freedom and clearly favors a collective rather than a  single particle picture
of them.

These simple expectations are borne out in an explicit model calculation.
We shall employ the Flux Tube Model of Isgur and Paton \cite{IP} for
this purpose.
The model is extracted from the strong coupling limit of the QCD lattice 
Hamiltonian. The authors first split the Hamiltonian into blocks of distinct
``topologies" (in reference to the possible gauge invariant flux tube 
configurations)  and then make adiabatic and  small oscillation approximations
of the flux tube dynamics to arrive at an N-body discrete string-like model 
Hamiltonian
for gluonic degrees of freedom. This is meant to be operative at intermediate 
scales $a \sim b^{-1/2}$ where the strong coupling is order unity.
The lattice spacing is denoted by $a$ and there are $N$ 
``beads'' (or links) evenly spaced between the $Q \bar Q$ pair. 
These considerations led Isgur and Paton to write the following 
model Hamiltonian

\begin{equation}
H_{FT} = b_0 r + \sum_{n=1}^{N} {1\over 2 b_0 a} {\bf p}^2_n + {b_0 a \over 2}
\sum_{n=1}^{N+1} (\bbox{\chi}(n) - \bbox{\chi}(n-1))^2
\end{equation}

\noindent
Here $b_0$ is the bare string tension,  
${\bf p}_n$ is the conjugate momentum, and   
$\bbox{\chi}(n)$ is the transverse 
displacement vector at site $n$.  It has been generalized to include color
degrees of freedom: $\bbox{\chi} = \chi^c_{\lambda}(n)$, $c=1\cdots8$, 
$n=1\cdots N$, and $\lambda=1,2$.
The ends of the string are fixed at $n=0$ and $n=N+1$ by the static quark 
and antiquark positions and $r = a(N+1)$.

In normal coordinates the Flux Tube Hamiltonian is

\begin{equation}
H_{FT} = b_0 r + \sum_{n=1}^N\sum_{\lambda=1}^2 \left({1 \over 2 b_0 a} 
\pi_{n\lambda}^2 + b_0 a \omega_n^2 q_{n\lambda}^2 \right) \label{fto}
\end{equation}

\noindent
where the normal mode frequencies are given by 
$\omega_n = {2\over a} \sin({n\pi\over 2(N+1)})$. 
This Hamiltonian may be trivially diagonalized with the following 
canonical transformation:

\begin{equation}
\alpha^a_{n\lambda} = \sqrt{b_0 a \omega_n \over 2} q^a_{n \lambda} + {i \pi^a_{n\lambda}
\over \sqrt{2 b_0 a \omega_n}},
\end{equation}

\noindent
and one obtains

\begin{equation}
H_{FT} = \sum_{n\lambda} \omega_n \alpha_{n\lambda}^\dagger \alpha_{n\lambda}
+ b r - {1\over a} - {\pi \over 12 r} + \ldots
\end{equation}
The string tension has been renormalized by the zero point energy which also 
introduces a (divergent) constant and terms higher order in $1/r$ as shown above.

Evaluating the spin dependent potentials in the Flux Tube Model requires explicit
expressions for the electric and magnetic fields in a flux tube. It is therefore
necessary to extend the Flux Tube Model somewhat. We shall find it convenient
to work in an ``intermediate" coupling regime where we shall use the strong 
coupling lattice Hamiltonian for intuition and the weak coupling limit for
the identification of the electric and magnetic fields.
In the strong coupling limit the electric field operator simply counts links. 
It is therefore natural to map it onto 
a link displacement,

\begin{equation}
E^a_\lambda(n) \sim {\kappa \over a^3}( \chi^a_\lambda(n+1) - \chi^a_\lambda(n)). \label{ef}
\end{equation}

\noindent
For the sake of clarity, we shall take $\r_q = {\bf 0}$ and $\r_{\bar q} = 
r \hat {\bf z}$ from now on. This implies that $\lambda = x,y$. The factor $\kappa$ is
an arbitrary constant and will be identified later.

The commutation relation between the electric and magnetic fields is

\begin{equation}
[E^a_i(\x), B^b_j(\y)] = i \epsilon^{ijk} \nabla^k_y \delta(\x-\y) \delta^{ab} +
{\cal O}(g), \label{com}
\end{equation}

\noindent
which implies that the magnetic field operator may be defined as the momentum conjugate 
to $\chi_\lambda$~\cite{MP} 

\begin{equation}
B^a_\lambda(n) = {-i \over \kappa a} {\partial \over \partial 
\chi^a_{\bar \lambda}(n)}. \label{bf}
\end{equation}

\noindent
The $\epsilon^{ijk}$ of the commutator in Eq.~(\ref{com}) is taken into account by the index 
$\bar \lambda$,  $\chi_{\bar x}(n) = - \chi_y(n)$ and 
$\chi_{\bar y}(n)= \chi_x(n)$. 
Note that this relationship is physically sensible because the magnetic 
field operator maps onto the plaquette in the strong coupling limit and 
the application of a 
plaquette to a flux tube has the effect of 
moving a link one unit in the one of the directions transverse to the $\bar Q Q$ axis 
 with the magnetic field pointing in the other transverse direction.

Substituting Eqs.~(\ref{ef}) and ~(\ref{bf}) into Eq.~(29)
and setting $\kappa = 
a\sqrt{b_0}$ yields

\begin{equation}
H_{FT} = b_0 r + {1\over 2}\int d\x \left((E_\perp^a(\x))^2 + 
(B_\perp^a(\x))^2\right).
\end{equation}

\noindent
This is reminiscent of the Coulomb gauge Hamiltonian where $H_{YM}$ involves
transverse fields only and where the nonabelian Coulomb interaction has been
replaced with a linear potential.
Thus a satisfying
consistency has been achieved with this approach.

It is now a simple matter to write the field operators in terms of the collective 
phonon operators, $\alpha$. 
One obtains 

\begin{equation}
E^a_\lambda(\x_n,t)= {\kappa \over a^3 \sqrt{b_0 r}} \sum_m \left(
\sin({m\pi \over N+1} (n+1)) - \sin({m\pi\over N+1}n)\right) {1 \over 
\sqrt{\omega_m}} \left(
\alpha_{m\lambda}^a \e^{-i\omega_m t} + \alpha_{m\lambda}^{a\dagger}
\e^{+i\omega_m t}\right) 
\end{equation}

\noindent
and

\begin{equation}
B^a_\lambda(\x_n,t) = {-i\over \kappa} \sqrt{b_0\over r} \sum_m 
\sin({m\pi \over N+1}n) 
\sqrt{\omega_m} \left( \alpha_{m\lambda}^a \e^{-i\omega_m t} - 
\alpha_{m\lambda}^{a\dagger} \e^{+i\omega_m t}\right) 
\end{equation}

Substituting these expressions into Eqs.~(\ref{v1ef}) and ~(\ref{v2ef}), 
taking the phonon operator matrix elements in the 0-phonon number ${\bar Q} Q$ ground state,
doing the time integrals, and evaluating the sum over modes yields

\begin{equation}
V_1(r) = -{g^2 \over 2 a^2} C_F r
\end{equation}

\noindent
and 

\begin{equation}
V_2(r) = \lim_{N\rightarrow \infty} {g^2 \over 2 a^2} C_F {r \over N}.
\end{equation}

\noindent
In the strong coupling limit one has $b = g^2/(2 a^2) C_F$ so that the anticipated
expression for $V_1$ emerges in a natural way.  
Furthermore, $V_2$ approaches zero like $1/N$; this is
also true for $V_3$ and $V_4$. The latter point is illustrated in Fig. 1 where the 
correlation of electric and magnetic fields versus separation along the 
flux tube is shown. As expected the fields become completely decorrelated as
the number of intervening degrees of freedom becomes large. Notice that this 
implies
that a constituent gluon model of hybrids would not have been able to produce
an effective scalar interaction. 

\hbox to \hsize{%
\begin{minipage}[t]{\hsize}
\begin{figure}
\epsfxsize=4in
\hbox to \hsize{\hss\epsffile{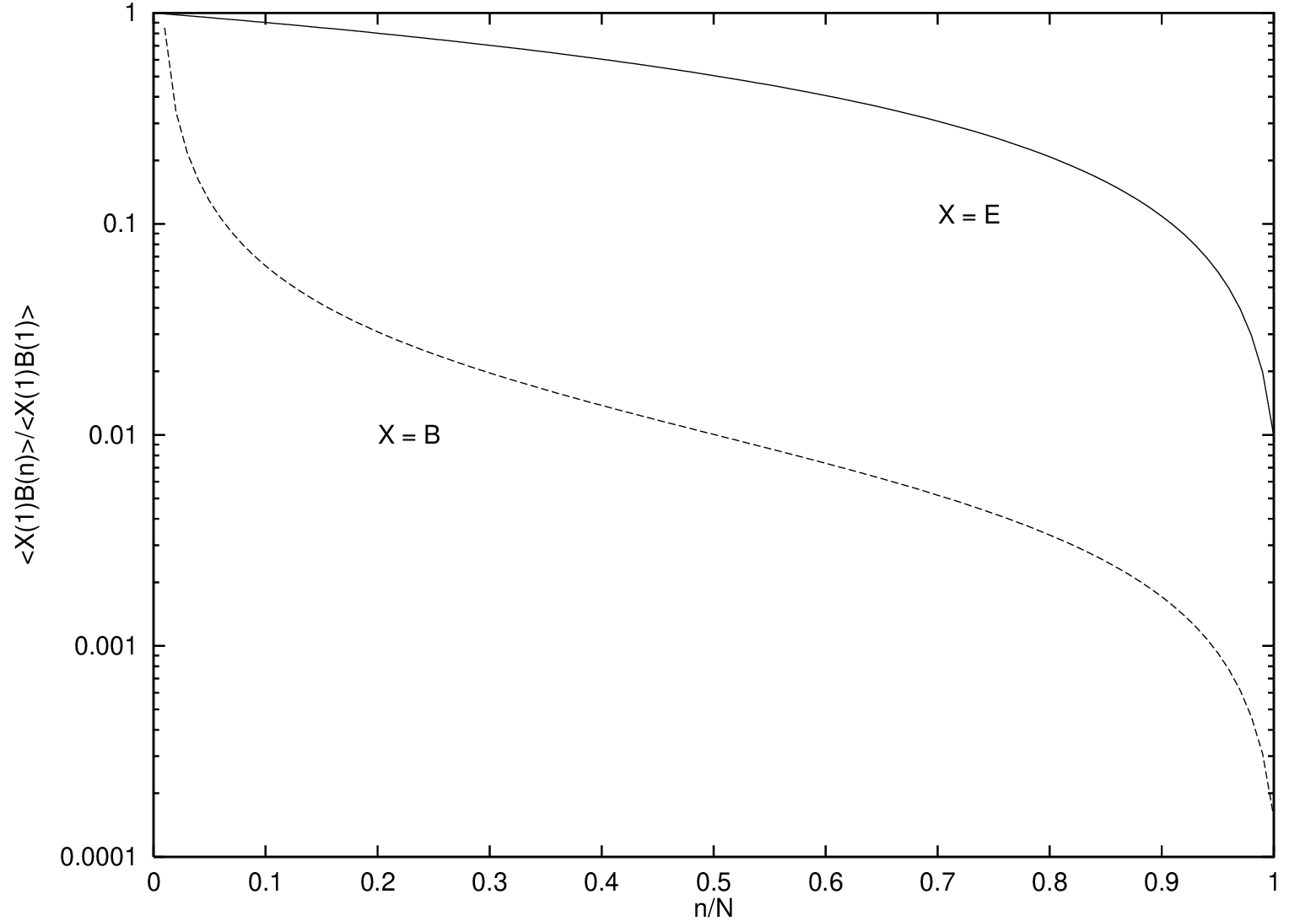}\hss}
\end{figure}
\end{minipage}}
\begin{center}
  {\small Fig.~1. Field Correlation Functions (for N=100).}
\end{center}

\section{Conclusions}

Spin splittings and lattice calculations indicate that confinement is scalar in
nature. This conflicts with many relativistic models of QCD which require vector
confinement. For example, a chirally symmetric interaction is needed if 
pseudo-Goldstone pions  and spontaneous chiral symmetry breaking  
 are to be generated dynamically.  
 Furthermore it appears to be impossible
to build a stable vacuum with a scalar kernel. We have examined
this issue with the heavy quark limit of the Coulomb gauge QCD Hamiltonian. 
This approach is physically intuitive and is simpler to interpret and implement than
methods based on the Wilson loop. We found that the static confinement potential
must indeed be a Dirac timelike vector. Effective scalar interactions are generated at
order $1/m^2$ by nonperturbative mixing with hybrid states. 

We have argued that the long range spin-spin ($V_3$ and $V_4$) and the vector-like
spin-orbit potentials ($V_2$) should all be zero since they involve field 
correlation functions evaluated between quark and antiquark. This statement
follows by assuming that the gluonic degrees of freedom collapse into a 
flux tube-like configuration, as shown by lattice gauge theory. Alternatively, the scalar-like
spin-orbit potential ($V_1$) is proportional to the matrix element of the electric
and magnetic fields evaluated at the same point and hence is expected to be 
nonzero. Explicit calculations of the relevant matrix elements were carried out in 
the Flux Tube Model. The model was extended to include color degrees of
freedom and to map the chromoelectric and chromomagnetic fields to flux tube 
phonon operators. The results obtained were in agreement with our general 
arguments and with Gromes' relation. 

A consistent picture of
the Dirac structure of confinement has emerged. The static central potential is
timelike vector while the spin-dependent structure mimics the nonrelativistic
reduction of an effective scalar interaction. This implies that it is incorrect
to employ a scalar confinement kernel when doing calculations with light quarks.
Note however that it would be acceptable to use scalar confinement when working 
explicitly in the chiral symmetry broken phase, {\it i.e.}, 
with constituent quarks in the nonrelativistic limit. The work presented here also
implies that a constituent gluon picture of hybrids will yield incorrect results
for certain observables. For example, $V_1$ and $V_2$ would be of comparable
magnitude in a constituent gluon model. In general, these types of models
must fail when 
nonlocal properties of the gluonic configuration are considered. Alternatively, 
it is possible that they
perform quite well when evaluating global properties of gluonics such as 
the hybrid spectrum. 

The application of these results to spin splittings in 
heavy quarkonia 
is not straightforward. For example, there is the possibility of large 
${\cal O}(1/m^3)$ corrections to the splittings. Light quark loop effects may 
also contribute to spin-dependent forces. It is, unfortunately, rather difficult
to quantify these effects. Perhaps the best hope is with high precision NRQCD 
lattice calculations. Light quark loops may be studied by examining shifts in 
the predicted spin splittings between quenched and unquenched calculations. One
naively expects that these effects will be independent of the (heavy) valence 
quark mass. Alternatively, the higher order corrections should become larger as
one moves from the $\Upsilon$ system to the $J/\psi$ system (the quenched
approximation may used to test this). Lattice results for the quenched 
$\Upsilon$ spectrum\cite{NRQCD1}
appear to be very close to experiment -- supporting the idea that light quark loop
effects are negligible (or at least may be absorbed into the parameters). Indeed, 
the calculated value of the ratio of P-wave splittings
$\rho = [M(^3P_2)-M(^3P_1)]/[M(^3P_1) - M(^3P_0)]$ yields 0.7(3)\cite{NRQCD2}, in
good agreement with the experimental value of 0.66(2). Unfortunately, this 
may not be regarded as definitive since $\rho$ was evaluated at $\beta = 6/g^2 = 
6.0$; a similar calculation at $\beta = 5.7$ gives $\rho = 1.4(4)$, very far from the
experimental ratio.  The situation in the $J/\psi$ spectrum is somewhat worse. 
There, Davies {\it et al.}\cite{NRQCD2} find $\rho = 1.2(2)$, while the experimental
value is 0.48(1). Thus it appears that higher order corrections in the inverse quark mass
may become substantial at the charm quark scale. Clearly further lattice work
is required to settle this issue. The effects of unquenching may also be
studied theoretically (although this will necessarily be model dependent). For
example, Eichten {\it et al.}\cite{cornell} have used the Wigner-Weisskopf method
in a vector confinement model to study the importance of virtual meson decays.
Geiger and Isgur\cite{GI} have also studied such effects on the structure of the
nucleon using the $^3P_0$ model to incorporate meson loops.

The methodology which we have adopted here -- using the heavy quark expansion of
the Coulomb gauge QCD Hamiltonian to identify pertinent matrix elements and then
using the (extended) Flux Tube Model to calculate these -- is potentially very
useful. For example, we may calculate the spin independent shifts to the static
potential in precisely the same manner. This should prove interesting since we 
expect that they may not be obtained through the nonrelativistic reduction of
an assumed kernel. Rather we will see for the first time evidence of the 
dynamical nature of confinement in the effective $1/m^2$ structure. Spin-orbit
splittings in baryons have been something of a mystery for a long time\cite{IK}. 
These should be accessible with the same techniques used here -- assuming that one
can find a satisfactory quantum flux tube model of gluonic excitations in baryons. It
should also be possible to examine the strong decays of heavy quarkonia with
this approach. This is of interest since these processes are 
poorly understood and are of central importance to many phenomena of current 
interest. Finally, most current models of hadrons do not contain dynamical 
(nonperturbative) gluonic degrees of freedom. It should, however, be possible
to include them in a way which is consistent with the Flux Tube Model. We are
currently investigating this possibility in the context of the Dynamical Quark 
Model\cite{ss}.

\acknowledgements
The authors are grateful to Nathan Isgur for fruitful discussions. 
ES acknowledges the financial support of the DOE under grant 
DE-FG02-96ER40944.

%\begin{figure}
%\caption{Field Correlation Functions (for N=100).}
%\label{fig1}
%\end{figure}

\end{document}